\documentclass[conference,usenatbib]{basi}
\usepackage[T1]{fontenc}
\usepackage[british]{babel}
\usepackage[varg]{txfonts}
\usepackage{amssymb} 
\usepackage{graphicx,color}
\usepackage{rotating}
\usepackage{dcolumn}
\usepackage{epstopdf}
\usepackage{epsfig}
\usepackage[a4paper=true]{hyperref}
\hypersetup{colorlinks = true, linkcolor = blue, anchorcolor = green, citecolor = blue, filecolor = blue, pagecolor = red, urlcolor = blue}
\begin{document}
\title[P. Rani \& C. S. Stalin]{Hard X-ray flux variations in AGN from NuSTAR}
\author[Priyanka Rani \& C. S. Stalin]%
       {Priyanka Rani\thanks{email: \texttt{priyanka@iiap.res.in}},
       C.~S.~ Stalin \\
       Indian Institute of Astrophysics, Block II, Koramangala, 
       Bangalore 560 034, India\\
       }

\pubyear{2015}
\volume{12}
\pagerange{\pageref{firstpage}--\pageref{lastpage}}

\date{Received --- ; accepted ---}

\maketitle
\label{firstpage}

\begin{abstract}
We present here our 
results on the hour like time scale X-ray flux variations
in a sample of active galactic nuclei using data from the 
{\it Nuclear Spectroscopic
Telescope Array (NuSTAR)}.  
We find that in the 3$-$79 keV band, 
BL Lacs are more variable than flat spectrum radio quasars and  
Seyfert galaxies. Among Seyferts, Seyfert 2s 
are more variable than Seyfert 1s.  
Also, radio-loud quasars are more variable
in the hard (10$-$79 keV) band than the soft (3$-$10 keV)  band while, 
Seyfert galaxies tend to show more
variations in the soft band relative to the hard band. 
\end{abstract}

\section{Introduction}\label{s:intro}

A distinctive property of active galactic nuclei (AGN)  
is that they show flux variations over the entire electromagnetic spectrum
over a wide range of timescales. 
Among the various wavelengths, hard X-rays are less affected by absorption, and thus, by exploring 
hard X-ray flux variability, one 
can put strong constraints on the physics in the innermost regions of AGN
that are not resolvable using any existing imaging facilities.  
At hard X-ray wavelengths, until recently, we have knowledge only on the 
long term variability properties of  AGN, that too for a limited number of 
sources   
mainly from Swift/BAT and RXTE/PCA observations. With the availability of the hard X-ray focussing telescope,  
NuSTAR \citep{2013ApJ...770..103H} since June 2012, it is now possible to
 probe the hard X-ray flux variations in AGN on  a wide range of time scales.  
Here, we have carried out a systematic study on the hard X-ray variability
properties of a sample of AGN on hour like time scales using observations from NuSTAR.

\section{Sample, data reduction and analysis}
Our sample consists of 71 AGN, which includes 
4  BL Lac 
objects (BL Lacs), 3 flat spectrum radio quasars (FSRQs),  24 Seyfert 1 
galaxies and 40 Seyfert 2 galaxies.  
Data have been processed 
using the {\it NuSTAR} Data Analysis Software ({\tt NuSTARDAS})
v.1.4.1, following standard procedures and using the CALDB 
version 20141107. A circular region of $60''$ was taken to extract the source 
and background counts on the same detector.
Light curves were generated in the  soft (3$-$10 keV),  
hard (10$-$79 keV) and total (3$-$79 keV) X-ray bands. A sample 
light curve is shown in Fig. 1. To characterise flux variations, we have 
used the normalized excess variance ($F_{var}$) 
following \cite{2003MNRAS.345.1271V}.
Distribution of $F_{var}$ for the different types of sources 
in the 3$-$79 keV band is shown in Fig. 1.
\section{Results}
Our conclusions are  (i) all source in our sample showed X-ray variations,  
(ii) radio 
loud sources are  more variable in the hard band compared to the soft band
while Seyfert galaxies vary more in the soft band, (iii) among radio-loud 
sources, BL Lacs are more variable than FSRQs both in soft and hard
bands and (iv) among Seyferts, Seyfert 2s
are more variable than Seyfert 1s in both soft and 
hard X-ray bands.  

\begin{figure}
\begin{center}
\vspace*{-0.2cm}
\includegraphics[height=7cm,width=11cm]{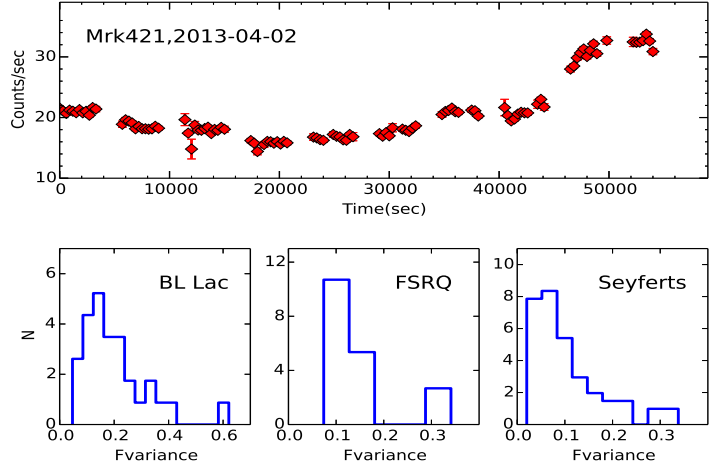}
\end{center}
\vspace*{-0.6cm}
\caption{A sample light curve for Mrk 421 (top) and the distribution of F$_{var}$ (bottom)}
\label{fig1}
\end{figure}



\end{document}